\documentclass[AMA,STIX1COL]{WileyNJD-v2}

\articletype{Article Type}%

\received{26 April 2016}
\revised{6 June 2016}
\accepted{6 June 2016}

\begin{document}

\title{Robust joint modelling of longitudinal and 
survival data with a time-varying degrees-of-freedom 
parameter}

\author[1]{Lisa M. McFetridge$\dagger$}

\author[2]{{\"O}zg{\"u}r Asar*$\dagger$}

\author[3]{Jonas Wallin}

\authormark{McFetridge \textsc{et al}}

\address[1]{\orgdiv{Mathematical Sciences Research Centre, School of Mathematics and Physics}, \orgname{Queen's University of Belfast}, \orgaddress{\state{Belfast}, \country{United Kingdom}}}

\address[2]{\orgdiv{Department of Biostatistics and Medical Informatics, Faculty of Medicine}, \orgname{Ac{\i}badem Mehmet Ali Ayd{\i}nlar University}, \orgaddress{\state{\.{I}stanbul}, \country{Turkey}}}

\address[3]{\orgdiv{Department of Statistics}, \orgname{Lund University}, \orgaddress{\state{Lund}, \country{Sweden}}}

\corres{*{\"O}zg{\"u}r Asar, Department of Biostatistics and Medical Informatics, Faculty of Medicine, Ac{\i}badem Mehmet Ali Ayd{\i}nlar University, \.{I}stanbul, Turkey. \email{ozgur.asar@acibadem.edu.tr | ozgurasarstat@gmail.com}}

%\presentaddress{This is sample for present address text this is sample for present address text}

\abstract[Summary]{Repeated measures of biomarkers have the potential of 
explaining hazards of survival outcomes. 
In practice, these measurements 
are intermittently measured and are known to be 
subject to 
substantial measurement error. 
Joint modelling of longitudinal and 
survival data enables us to associate intermittently measured 
error-prone biomarkers with risks of survival outcomes. 
Most of the joint models available in the literature have 
been built on the Gaussian assumption. 
This makes them sensitive to outliers. 
In this work, we study 
a range of robust models to address this issue. 
For medical data, it has been observed that outliers 
might occur 
with different frequencies over time. 
To address this, a new model with a time varying robustness is introduced. Through both a simulation study and analysis of two 
real-life data examples, this research not only stresses the need to account for longitudinal outliers in joint modelling research but also highlights the bias and inefficiency 
from not properly estimating the degrees-of-freedom parameter. 
Each technique presented in this work can be fitted using the {\tt R} package {\tt robjm}.}


%\jnlcitation{\cname{%
%\author{Williams K.}, 
%\author{B. Hoskins}, 
%\author{R. Lee}, 
%\author{G. Masato}, and 
%\author{T. Woollings}} (\cyear{2016}), 
%\ctitle{A regime analysis of Atlantic winter jet variability applied to evaluate HadGEM3-GC2}, \cjournal{Q.J.R. Meteorol. Soc.}, \cvol{2017;00:1--6}.}

\maketitle

%\footnotetext{{$^\dagger$} L.M. McFetridge and Ö. Asar are equal contributing authors.}
\makeatletter{\renewcommand*{\@makefnmark}{}
\footnotetext{{$^\dagger$} L.M. McFetridge and Ö. Asar are equal contributing authors.}\makeatother}

\section{Introduction}
\label{Introduction}

Biomarkers are used as proxies of one's health, 
 and are known to be subject to substantial measurement 
 error, due to biological and non-biological sources. 
 In prospective studies, the biomarkers are  
 repeatedly measured. These measurements 
 are made at intermittent time-points in practice rather 
 than continuously, and   
 times elapsed between successive measurements are typically unequal.
 It is of scientific interest to explain risks of 
 survival outcomes by repeated measures of biomarker data. 
 Two main obstacles are the intermittent nature of the 
 biomarker data and inherent measurement error. 
 Joint modelling of longitudinal and survival outcomes 
 enables us to associate repeated measures of biomarkers 
 with the risks of survival events, while taking into account 
 the aforementioned obstacles. 
 The framework typically combines two sub-models, 
 a mixed-effects model for repeated measures, 
 and a Cox model for survival data. The 
 sub-models are linked with shared parameters. 
 Reviews of relevant literature can be 
 found in 
 \citet{Tsiatis2004}, \citet{Diggle2008}, 
 \citet{Rizopoulos2012}, \citet{McCrink2013}, 
 \citet{asar2015} and \citet{Elashoff2017}. 

A prevailing assumption in the literature is that both random-effects and error terms in the mixed-effects sub-model follow Normal distributions. However, in most real-life problems, the 
data-sets include outliers for which the Normal assumption might 
be inadequate. Specifically, 
in prospective studies, 
two types of outliers may be present:
\begin{enumerate}
\item[i)] \textbf{b-outliers}: Outlying individuals within the population that do not conform to population trends. These are outliers within the longitudinal random-effects.
\item[ii)] \textbf{e-outliers}: Outlying observations within an individual's set of measurements that do not follow the individual's own trend over time. These are outliers within the longitudinal random error.
\end{enumerate}

Only in recent years have studies been undertaken to investigate the negative impacts that the presence of longitudinal outliers under the Normal assumption can have~\cite{Li2009, Huang2010, baghfalaki2013, McCrink2014, asar2019}. 
Robust joint models replace the Normality assumption with t-distributional assumptions for the random terms. In doing so, the heavier tails down-weigh the detrimental impact of longitudinal outliers within the joint modelling framework. 
Initial research on robust joint modelling considered only the $e$-outliers~\citep{Li2009, Huang2010, Taylor2013}, and both $e$- and $b$-outliers~\citep{McCrink2014, asar2019, Song2012, Baghfalaki2014}. Each of these studies found that the parameter estimates and corresponding standard errors are sensitive to the Normality assumptions when outliers are present, with t-distributional assumptions alleviating such bias. If utilised when outliers are not present, unbiased estimates are obtained, though~\citet{Li2009} and~\citet{Huang2010} noted that the robust joint models give slightly higher standard errors for the longitudinal parameters. 
This initial work however restrictively fixes the degrees-of-freedom at a constant chosen by the user, an assumption alleviated by~\citet{baghfalaki2013}, \citet{asar2019} and \citet{Baghfalaki2014}, who utilised Bayesian approaches, and~\citet{McCrink2014}, who utilised a frequentist approach, to allow estimation of the degrees-of-freedom to be dictated by the data. With estimated degrees-of-freedom, the t-distributional assumptions of the robust joint model will approximate normality in the absence of outliers.

All of the aforementioned research, however, assumes that the degrees-of-freedom is constant, unchanging over time, an assumption prevalent in the literature for robust mixed models alongside their robust joint model counterparts. To address this issue, we propose a robust joint model which can account for the situation where the impact of outliers can change over time. Varying frequency of outliers is a very likely scenario, for example, when patients are given new treatments, they typically take time to stabilise and adjust to these treatments. This period of adjustment means that they are more prone to demonstrating responses which outlie from the expected and thus the patients' measurements may not be consistent with the population average and change over time as they grow accustomed to the new treatments. Incorrectly modelling such scenarios limits our ability to fully decipher the relationship between how patients’ responses change over time and the impact this has on their risk of an adverse event.

This work explores two examples, one on dialysis, 
and another on liver cirrhosis. 
The first example analyses the haemoglobin levels of Northern Ireland (NI) renal patients after commencing haemodialysis treatment. The second example, utilising the Mayo Clinic Primary Biliary Cirrhosis data-set, analyses data obtained from patients with primary biliary cirrhosis who participated in a randomized placebo controlled trial of the drug D-penicillamine. In both case studies, it is reasonable to assume that patients may take time to adjust to the new treatments and thus would be more prone to outlying responses.

A more detailed discussion on the two motivational examples is given in Section~\ref{MotivatingExamples}. The rest of this paper is organised as follows. In Section~\ref{Methodology}, the methodology for the various joint modelling approaches investigated in the paper is presented, including the introduction of a robust joint model with time-varying degrees-of-freedom. Section~\ref{SimulationStudy} presents a simulation study comparing the approaches assuming an underlying structure of time-varying degrees-of-freedom. Analysis of the two motivating examples is presented in Section~\ref{DataAnalysis} followed by discussions in Section~\ref{Discussion}.

\section{Motivating Examples}
\label{MotivatingExamples}

\subsection{NI Renal data}
\label{MotivatingExamples:NIRenal}

Recent renal research has shown the potential for patients' haemoglobin (Hb) levels as an emerging biomarker for chronic kidney disease \citep{Chen2015}. It is known within the renal research that in the initial stages after commencing haemodialysis, patients' haemoglobin levels tend to fluctuate to a large degree and is deterimental to their survival. This research aims to capture and model these initial stages of variability where there is a greater potential for patients to either have outlying observations from their own trends over time or be an outlier themselves from the population.

This work will analyse data that has been collected from seven different renal centres by the NI Renal Information Service. The NI renal data-set contains longitudinal information on 1,339 patients who have undergone haemodialysis treatment between April 2002 and July 2011~\citep{McCrink2013, McCrink2014, Donnelly2017}. In total, 27,064 Hb measurements were recorded, with a median of 14 observations per individual. Other health related measurements, including creatinine and urea levels, were additionally monitored. Individuals were observed for a minimum of 2 days, and a maximum of approximately 9 years throughout the study follow-up period. 44\% of the patients died during the study period with the remaining patients being right-censored. 

\begin{figure}[!ht]
\centering
  \includegraphics[scale = 0.6]{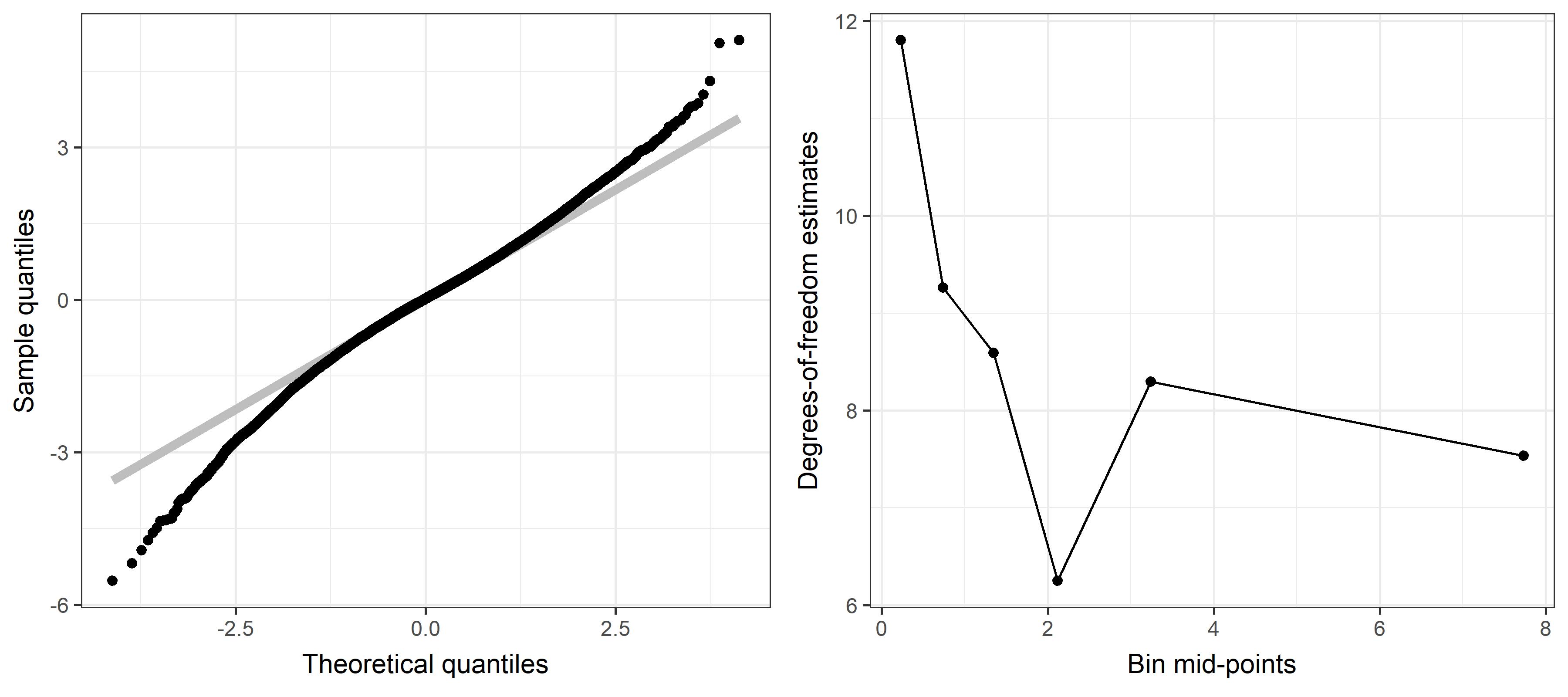}
  \caption{Plots to inspect tail behavior of the standardised 
  conditional residuals obtained from the joint model with Gaussian 
  assumptions for the renal dataset. Details of calculations are provided within Section~\ref{MotivatingExamples:NIRenal}. Left: Quantile-quantile plot 
  against standard Normal distribution; Right: Degrees-of-freedom parameter estimates.} 
  \label{fig:renal_exploratory_plots1}
\end{figure}

\begin{figure}[!ht]
\centering
  \includegraphics[scale = 0.6]{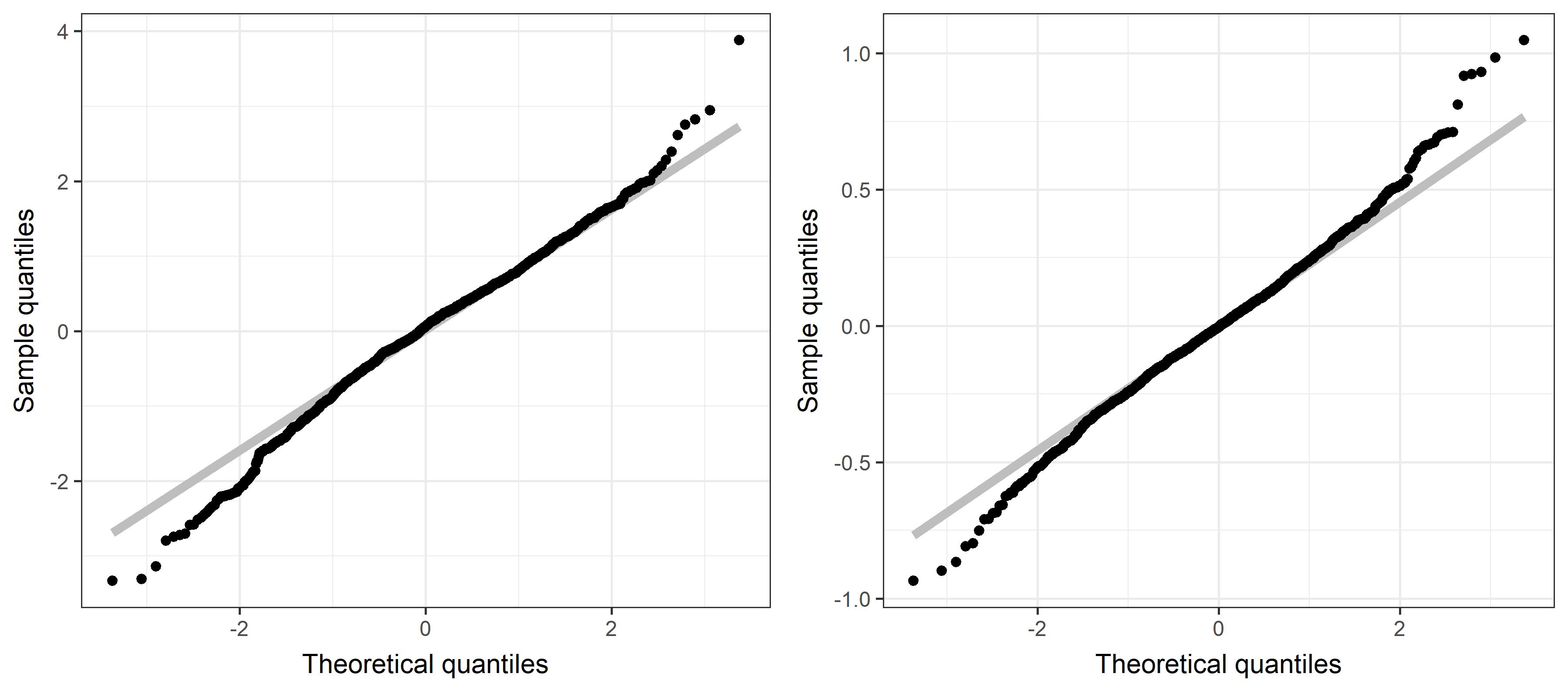}
  \caption{Plots to inspect the distributions of 
  the random-intercept (left) and random-slope (right) 
  terms against standard Normal based on the renal data.}
  \label{fig:renal_exploratory_plots2}
\end{figure}

Figures~\ref{fig:renal_exploratory_plots1} and~\ref{fig:renal_exploratory_plots2} show the tail behaviour of the standardised conditional residuals and random-effects, respectively, based on a joint model fitted for the NI renal data with Gaussian assumptions for both random-effects and error terms. Degrees-of-freedom estimates for Figure~\ref{fig:renal_exploratory_plots1} were obtained by fitting a univariate $t$ distribution such that $t(\mu, \sigma, \nu)$, where $\mu$ is the location parameter, $\sigma$ the scale, $\nu$ the degrees-of-freedom. Time (in years) was divided into bins based on 20\% percentiles of the follow-up time. As illustrated by Figures~\ref{fig:renal_exploratory_plots1} and~\ref{fig:renal_exploratory_plots2}, there are departures from Normality for both random-effects and error terms. Figure~\ref{fig:renal_exploratory_plots1} further indicates that the degrees-of-freedom parameter of the $t$-distribution for the residuals varies over time, an anticipated result based on the known impact on patients' Hb levels initially as they adjust to the commencement of haemodialysis treatment. Such changes over time will be accounted for through the time-varying degrees-of-freedom formulation introduced in Section~\ref{sec:mod4}.

\subsection{Primary biliary cirrhosis (PBC) data}
\label{MotivatingExamples:PBC}

A second motivating example arises from data 
collected from 312 patients with primary biliary 
cirrhosis (PBC), a rare autoimmune disease which 
results in cirrhosis of the liver and fatality. 
Patients suffering from PBC produce increased 
levels of alkaline in the blood. The PBC data-set 
was collected from the Mayo Clinic trial conducted between 
1974 and 1984, which aimed 
to assess the effectiveness of the D-penicillamine 
drug \citep{Murtaugh1994}. Of the 312 patients, 
158 were randomly assigned 
to the drug, and 154 to placebo. Median baseline age was 
48.8 (minimum = 26.28, maximum = 78.44). 
In total, 1,885 repeated 
alkaline observations were collected over the 10 year 
trial period, with a median of 5 (minimum = 1, maximum = 16) observations per individual. Individuals were observed for 
a minimum of 0.1 years, and a maximum of 14.3 years. 
140 (45\%) patients died during the study-period. 

\begin{figure}[t]
\centering
  \includegraphics[scale = 0.6]{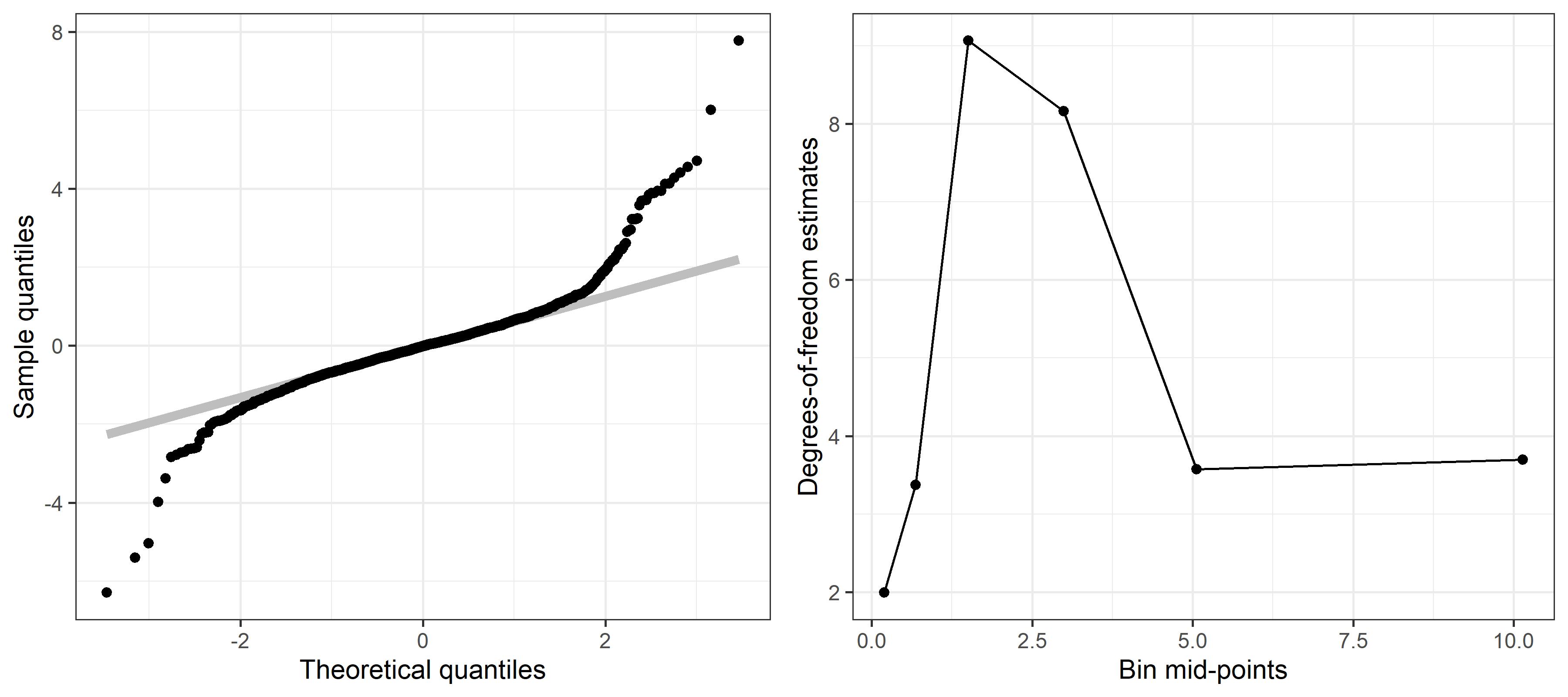}
  \caption{Plots to inspect the standardised 
  conditional residuals obtained from the joint model with Gaussian 
  assumptions for the PBC data-set. 
  Left: Quantile-quantile plot 
  against standard Normal distribution, 
  Right: degrees-of-freedom parameter estimates. 
  Degrees-of-freedom are 
  obtained similar to Figure~\ref{fig:renal_exploratory_plots1}.}
  \label{fig:pbc_exploratory_plots}
\end{figure}

\begin{figure}[t]
\centering
  \includegraphics[scale = 0.6]{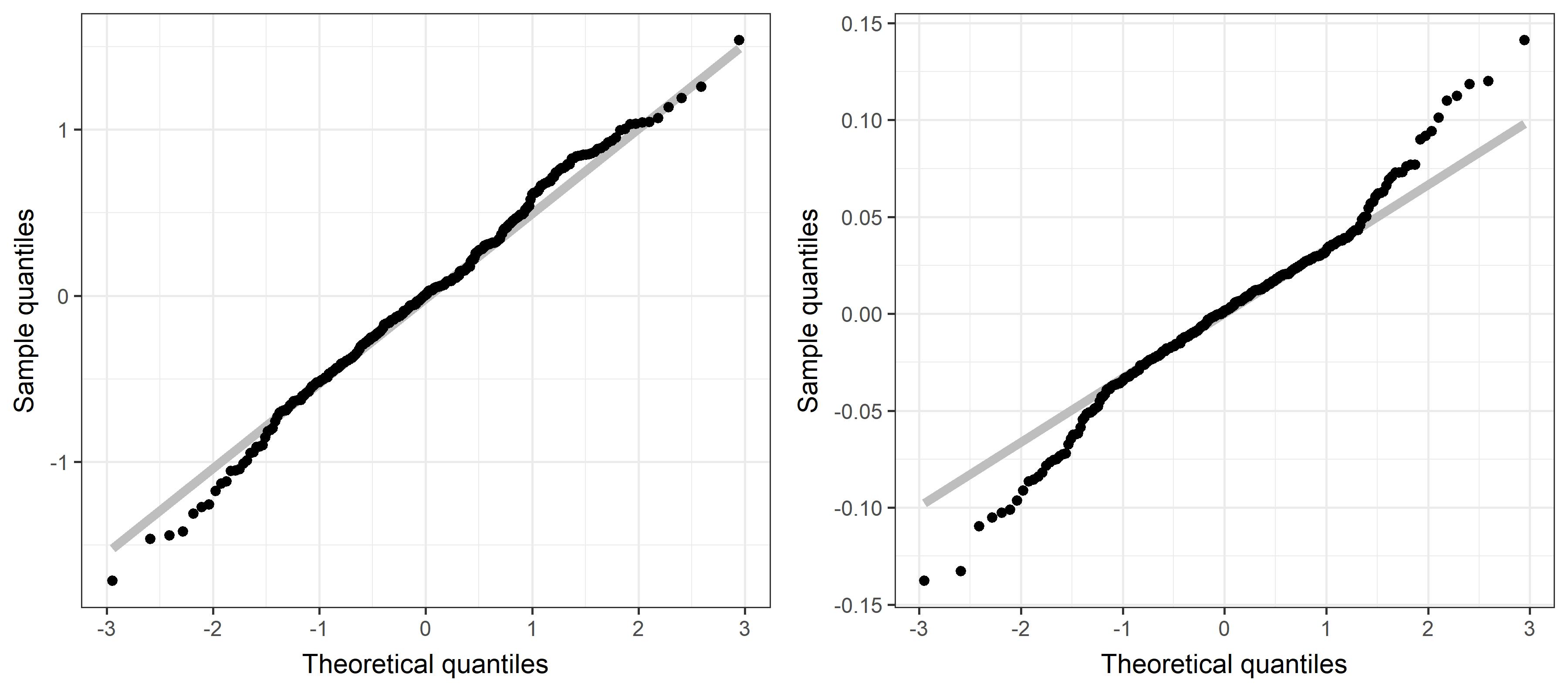}
  \caption{Plots to inspect the distributions of 
  the random-intercept (left) and random-slope (right) 
  terms against standard Normal.}
  \label{fig:pbc_exploratory_plots2}
\end{figure}

The quantile-quantile plot 
for the standardised 
conditional residuals under the joint model with 
Gaussian assumption (left panel of 
Figure \ref{fig:pbc_exploratory_plots}) 
indicates heavier tails than Gaussian. 
Estimated degrees-of-freedom
from univariate t distributions fitted to 
residuals that fall into bins based on 
20\% percentiles of follow-up time
indicate time-varying degrees-of-freedom 
structure up to some extent. 
Figure \ref{fig:pbc_exploratory_plots2} indicates that normality may be a reasonable assumption for the random effects.

\section{Approaches for Robust Joint Modelling}
\label{Methodology}

\subsection{Notation}
\label{sec:notation}
  
 Before introducing the models, we present the general notation used thoughout the paper.   
 Let $Y_{ij}$ denote the $j^{\text{th}}$ $(j = 1, \ldots, m_i)$ repeated measurement 
 belonging to subject $i$ $(i = 1, \ldots, n)$ collected at time $t_{ij}$,  
 $t_i = \{t_{i1}, \ldots, t_{im_i} \}$ the set of the follow-up times at which 
 $Y_{ij}$'s are collected, 
 $a_i = \{a_{i1} \ldots a_{1l}\}$ baseline covariate information, and 
 $S_i$ survival time. Here, $S_i$ is subject to right-censoring. 
 Hence, we introduce an additional random variable, $E_i$, 
 defined by $E_i = I(S_i^* \leq C_i)$, where $I(\cdot)$ being indicator 
 function, $S_i^*$ true survival time for subject $i$, 
 $C_i$ censoring time for the subject. 
 $C_i$ is defined as $C_i = \min(C, D_i)$, with 
 $C$ is the study end-time and $D_i$ is the drop-out time for subject $i$.
 
 \subsection{Joint modelling of longitudinal and survival data}
 \label{sec:gaussian_joint}
 
 The framework for the so-called 
 joint models for longitudinal 
 and survival data under the 
 shared-parameter paradigm is given by
 \begin{align}
 Y_{ij} &= Y_i^*(t_{ij}) + Z_{ij}, \nonumber\\ 
        &= {\boldsymbol x}_{ij}^{\top} {\boldsymbol \alpha} + {\boldsymbol d}_{ij}^{\top} {\boldsymbol B}_i + Z_{ij}, \\
 h_i(t) &= h_0(t) \exp \left( {\boldsymbol c}_i^{\top} {\boldsymbol \omega} + f(\mathcal{Y}_i^*(t); {\boldsymbol \eta}) \right).			
 \end{align}
 The framework allows for biomarker values (the longitudinal data) to be measured with error, i.e. the observed 
 data, $Y_{ij}$, is composed of underlying continuous-time 
 signal at time $t_{ij}$, $Y_i^*(t_{ij})$, and noise, $Z_{ij}$. The signal is de-composed into 
 fixed-effects, ${\boldsymbol x}_{ij}^{\top} {\boldsymbol \alpha}$, 
 and random-effects, ${\boldsymbol d}_{ij}^{\top} {\boldsymbol B}_i$. 
 Here, ${\boldsymbol x}_{ij}$ is a $p \times 1$ matrix structured by $a_i$ and $t_i$. 
 ${\boldsymbol \alpha}$ is a $p \times 1$ matrix of regression coefficients 
 as in multiple linear regression. 
 ${\boldsymbol d}_{ij}$ is a $q \times 1$ matrix, that is typically 
 a subset of ${\boldsymbol x}_{ij}$. 
 ${\boldsymbol B}_i$ are subject-specific coefficients 
 that take into account heterogeneity between subjects. 
 $h_i(t)$ is the hazard of survival event for subject $i$ 
 at time $t$. 
 $h_0(t)$ is the baseline hazard that 
 can be specified using hazard functions 
 of parametric distributions, e.g. Weibull; 
 left un-specified as in \citet{cox1972}; 
 assumed to be piece-wise constant; 
 or be expressed in terms of splines, 
 e.g. natural cubic or B-splines. 
 ${\boldsymbol c}_i$ is a $g \times 1$ matrix 
 with elements from $a_i$, 
 ${\boldsymbol \omega}$ $g \times 1$ vector of 
 regression coefficients. 
 $f(\mathcal{Y}_i^*(t); {\boldsymbol \eta})$ is the term  
 for taking into account the association 
 between hazard of survival event and 
 features of biomarker process, 
 with $f(\cdot)$ being a known function. 
 A popular choice is to use the 
 current value parametrisation such that 
 $f(\mathcal{Y}_i^*(t); {\boldsymbol \eta}) = \eta Y_i^*(t)$. 
 For other choices, see \citet{hickey2016}.  
 
 The classical model assumptions 
 for the random-effects and 
 error components are zero-mean 
 Gaussian distributions such that  
 \begin{align*}
 \boldsymbol{\mathrm{B}}_i &\sim \mathcal{N} \left( \boldsymbol{\mathrm{0}},\boldsymbol{\mathrm{\Sigma}} \right), \\
 Z_{ij} &\sim \mathcal{N} \left( 0, \sigma^2 \right),  \\
 \boldsymbol{\mathrm{B}}_i \perp Z_{ij}, \ \ &Z_{ij} \perp Z_{ij^{\prime}} \ \mbox{for} \ j \neq j^{\prime},
 \end{align*}
where $\boldsymbol{\mathrm{\Sigma}}$ is the covariance matrix 
of the random-effects, and $\sigma$ is the standard deviation of the measurement error. In what follows, 
we extend the assumptions beyond the Gaussian. 
  
 \subsection{Robust joint modelling}
 \label{sec:robust_joint}
 The robustness of the joint model 
 will be determined by the distributions 
 of the mixing variables, 
 $\boldsymbol{\mathrm{B}}_i$ and $Z_{ij}$. 
 The tail of the density for ${\boldsymbol{\mathrm{B}}}_i$ 
 determines robustness against $b$-outliers 
 and the tail of the density for $Z_{ij}$ 
 determines the robustness against $e$-outliers. 
We consider symmetric robust distributions for 
 both the random-effects and error components 
 through normal-variance mixtures such that
 \begin{align*}
 	\boldsymbol{\mathrm{B}}_i &= \boldsymbol{\mathrm{\Sigma}}^{1/2}\sqrt{V^B_i} \boldsymbol{\mathrm{B}}^*_i,  \\
 	Z_{ij} &= \sigma \sqrt{V^Z_{ij}}Z^*_{ij},  
 \end{align*}
 where $ \boldsymbol{\mathrm{B}}^*_i \sim \mathcal{N}(\boldsymbol{\mathrm{0}},\boldsymbol{\mathrm{I}}_{q\times q})$ and $Z^*_{ij} \sim \mathcal{N}(0,1)$. 
 This formulation is flexible and includes 
 widely used distributions as special 
 cases \citep{asar2018}. 
 The tail behaviour is 
 determined by the distribution of 
 $V_i^B$ and $V_{ij}^Z$. The special case $V_i^B = 1$ and $V_{ij}^Z = 1$ 
 recovers the Gaussian joint model (see Section \ref{sec:gaussian_joint}). 
   In this study, we specifically consider 
  inverse Gamma distribution $\mathcal{IG}$ 
  with equal shape and 
  scale parameters for the $V$ terms 
  which results a 
  t-distribution. 
 
Ideally, one can use the posterior distribution of $V_i^B$ or $V_{ij}^Z$ to detect outliers. If the posterior distribution of $V_i^B$ ($V_{ij}^Z$) is concentrated at large values this should indicate a $b$-outlier for subject $i$ ($e$-outlier for observation $j$ of patient $i$). However, in much of the robust modelling literature dependence between $V_i^B$ and $V_{ij}^Z$, 
e.g. see \cite{pinheiro2001}, and  
$V_{ij}^Z$ and $V_{ij^{\prime}}^Z$, 
e.g. \citet{baghfalaki2013}, 
have been introduced 
(mainly to simplify the inferential procedure). 
These dependencies, hovewer, make the 
aforementioned interpretation on outlier detection 
impossible.  

  In what follows, we will set three   
  robust joint model formulations 
  that are available in the literature 
  and discuss implications.
 
 \subsubsection{Approach 1}
 \citet{McCrink2014} considered 
  \begin{align*}
  V^B_i    &= V_i, \\
  V^Z_{ij} &= V_i, \\
  V_i \sim &\mathcal{IG}(\gamma/2, \gamma/2).
  \end{align*}
  The mixing variable is equal for both 
  random-effects and error components. 
  This implies that $b$- and $e$-outliers 
  must be present simultaneously for a subject, 
  which would be a strong assumption. 
  Note that under this assumption 
  ${\boldsymbol{\mathrm{B}}}_i$ and ${\boldsymbol{\mathrm{Z}}}_i = [Z_{i1} \ldots Z_{im_i}]^{\top}$ 
  will be jointly multivariate $t$ with 
  a single degrees-of-freedom parameter, 
  and the following 
  properties do not hold: 
  $\boldsymbol{\mathrm{B}}_i \perp Z_{ij}$, and 
  $Z_{ij} \perp Z_{ij^{\prime}}$ for $j \neq j^{\prime}$. 

 \subsubsection{Approach 2}
 \citet{baghfalaki2013} considered a robust joint model 
  by setting 
 \begin{align*}
 V^B_i &= V^B_i \sim \mathcal{IG}(\phi/2, \phi/2), \\
 V^Z_{ij} &= V^Z_i \sim \mathcal{IG}(\delta/2, \delta/2). 
 \end{align*}
 Under this approach, 
 $\boldsymbol{\mathrm{B}}_i$ and ${\boldsymbol{\mathrm{Z}}}_i$ will be multivariate $t$, 
 with separate degrees-of-freedom parameters;  
 the property of $\boldsymbol{\mathrm{B}}_i \perp Z_{ij}$ holds, 
 whereas $Z_{ij} \perp Z_{ij^{\prime}}$ for $j \neq j^{\prime}$ 
 does not. As $Z_{ij}$ and $Z_{ij^{\prime}}$ 
 for $j \neq j^{\prime}$ share a common 
 $V^Z_i$, this approach can be seen as a random-effects 
 approach on the variance of $Z$.  
 This approach forces all the elements 
 of $\{Z_{ij}: j = 1, \ldots, m_i\}$ 
 to be outliers.  
 
 \subsubsection{Approach 3}
 \citet{asar2019} considered the setting of
 \begin{align*}
 V^B_i &= V^B_i \sim \mathcal{IG}(\phi/2, \phi/2), \\
 V^Z_{ij} &= V^Z_{ij} \sim \mathcal{IG}(\delta/2, \delta/2), 
 \end{align*}
 for joint modelling. Under this approach the 
 dependence between $Z_{ij}$ and $Z_{ij^{\prime}}$ 
 that was present in Approach 2 has been removed. 
 This is the most natural approach compared to the previous two, 
 and will be used as a base for our time varying 
 degrees-of-freedom formulation.

 \subsection{Appoach 4: Time-varying degrees-of-freedom formulation}
 \label{sec:mod4}

 By building on Approach 3, 
 we consider time-varying 
 degrees-of-freedom 
 parameter, $\delta(t)$, for $Z_{ij}$. 
 In practice, 
 we need to discretise time, 
 i.e. $\delta(t) = \delta(t_{ij})$. 
 This approach considers 
 \begin{align*}
  V^Z_{ij}  &\sim \mathcal{IG}(\delta_{ij}/2, \delta_{ij}/2),\\
 \delta_{ij} &= \exp(\delta_0 + {\boldsymbol a}_{ij}^{\top}  {\boldsymbol \beta} ).
 \end{align*}
Here, ${\boldsymbol a}_{ij}$ is a $s \times 1$ to be specified  
 by natural cubic splines or B-splines,  
 ${\boldsymbol \beta}$ associated coefficients. 
 Note that one obtains time invariant degrees-of-freedom as $\delta_{ij} = \delta = \exp(\delta_0)$,
 when ${\boldsymbol \beta} = {\boldsymbol 0}$. 

 %=========================================================
 %=========================================================
 \subsection{Priors}
 \label{sec:priors}

 We only give details about the prior distributions 
 that are assigned to the parameters based on  
 Approach 4, 
 since Approaches 1-3 are just special cases. 
 We set weakly-informative prior distributions 
 for the parameters. 
 Elements of 
 $\alpha$ are assigned 
 zero-mean Cauchy prior with 
 scale of 5, 
 $\mathcal{C}(0, 5)$, 
 only $\alpha_0$ was given $\mathcal{C}(0, 20)$. 
 ${\boldsymbol{\mathrm{\Sigma}}}$ is re-written as 
 ${\boldsymbol{\mathrm{R}}} {\boldsymbol{\mathrm{\Omega}}} {\boldsymbol{\mathrm{R}}}$, 
 with ${\boldsymbol{\mathrm{R}}}$ being a diagonal matrix 
 of scale parameters of $B_{h}$ $(h = 1, \ldots, q)$, 
 and ${\boldsymbol{\mathrm{\Omega}}}$ a correlation matrix \citep{stanmanual}. 
 Elements of ${\boldsymbol{\mathrm{R}}}$ are given 
 half-Cauchy, $\mathcal{C}_{+}(0, 5)$, 
 whereas elements of ${\boldsymbol{\mathrm{\Omega}}}$ 
 are given LKJ with the parameter of 2, $\mathcal{LKJ}(2)$. 
 Degrees-of-freedom parameter for ${\boldsymbol{\mathrm{B}}}_i$, 
 $\phi$, and 
 $\delta_0$ of $Z_{ij}$ 
 are given uniform priors, 
 between 2 and 100. 
 Elements of 
 ${\boldsymbol{\mathrm{\beta}}}$ 
 are given $\mathcal{C}(0, 5)$.  
 $\sigma$ is given $\mathcal{C}_{+}(0, 5)$. 
 Log-transformed elements of $h_0(t)$ and 
 elements of ${\boldsymbol{\mathrm{\omega}}}$ and ${\boldsymbol{\mathrm{\eta}}}$ 
 are given $\mathcal{C}(0, 5)$. 

 \subsection{Inference}
 \label{sec:inference}
 
 For inference, we use Markov Chain Monte Carlo methods to sample 
 from the joint posterior densities of the parameters and 
 latent variables given data. We specifically consider the 
 Hamiltonian Monte Carlo (HMC) \citep{neal2011} 
 as implemented in the No-U-Turn Sampler (NUTS) \citep{hoffman2014}. 
 We do not present the details of neither the
 likelihood function, 
 nor HMC and NUTS, 
 since, whereas the former is quite 
 straightforward (e.g. see \cite{asar2019}), 
 the second can be followed from the cited references. 
 
 Bespoke {\tt R} \citep{r2019} codes to fit the joint models 
 under Approaches 1-4 are available from the {\tt robjm} 
 package ({\tt https://github.com/ozgurasarstat/robjm}) that 
 internally uses the HMC sampling engine {\tt Stan} 
 \citep{carpenter2017} through the 
 {\tt R} package {\tt Rstan} \citep{rstan2018}. 

\section{Simulation study}
 \label{SimulationStudy}

A simulation study was conducted to investigate the effects of time-varying degrees-of-freedom in the estimation of parameters for the joint modelling approaches discussed in Section~\ref{Methodology}. A sample size of $n=250$ individuals was considered with 200 data-sets being simulated under the assumption of time-varying degrees-of-freedom for the longitudinal residuals. Note that 
we consider 200 as the number of replications mainly because 
of the computational cost. 

\begin{figure}[t]
\centering
  \includegraphics[scale = 0.65]{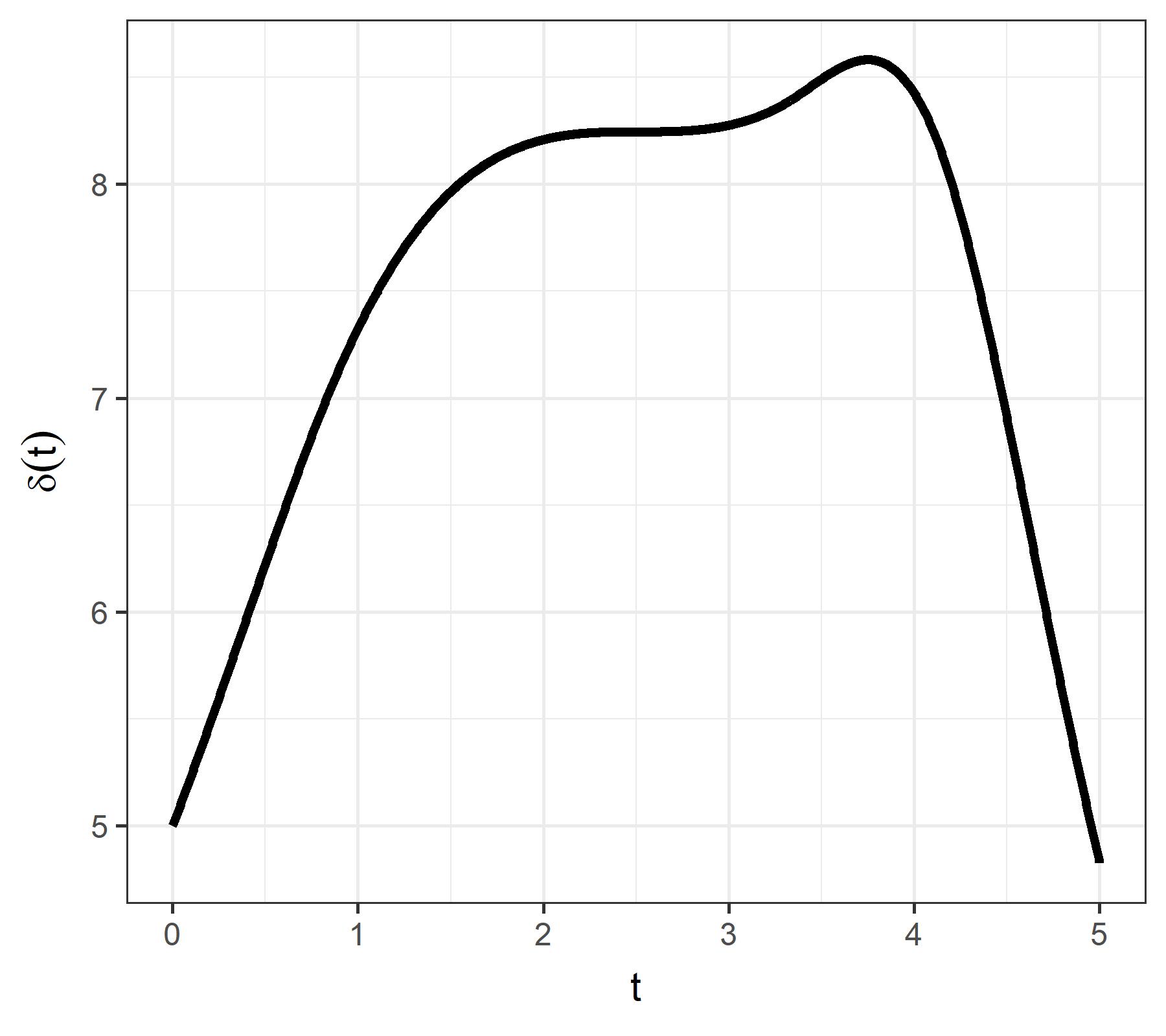}
  \caption{Underlying degrees-of-freedom assumption for simulation study.}
  \label{dofassumption}
\end{figure}

Data were generated under Approach 4. 
The assumed underlying structure for the time-varying degrees-of-freedom 
 is illustrated in Figure~\ref{dofassumption}. 
The longitudinal sub-model was given by: 
\begin{eqnarray}
Y_{ij} &=& Y^*_{i}(t_{ij}) + Z_{ij}, \nonumber \\
&=& \alpha_{1} + t_{ij}\alpha_{2} + B_{1i} + t_{ij}B_{2i} + Z_{ij},
\end{eqnarray}
where each individual has an average of 20 observations between time points 0 and 5. A random-intercept ($\alpha_{1} + B_{1i}$) and random-slope ($\alpha_2 + B_{2i}$) model was assumed to replicate what is most commonly used in the joint modelling literature. Survival data was generated from the following model:
\begin{eqnarray}
 h_i(t) &= h_0(t) \exp \left(X_i \omega + Y_i^*(t) \eta \right),	
\end{eqnarray}
where $X\sim\textrm{Bernoulli}(0.5)$, 
$h_0(t)$ specified by a Weibull baseline hazard, 
i.e. $h_0(t) = \lambda \nu t^{\nu - 1}$. 
There is a final truncation time of 5 after which non-informative right-censoring occurs. The true values for the unknown parameters are given in Table~\ref{SimTable}. 

Each robust model and the model with Gaussian 
assumptions 
for ${\bf B}_i$ and $Z_{ij}$, 
called the standard joint model,  
were estimated utilising the {\tt robjm} package 
with four chains each of length 2,000 with the first 1000 iterations considered as warm-up. The averages of posterior means, 
width of the corresponding 95\% credibility intervals (CI) and coverage (Cov.) of 200 replications for each model are collated in Table~\ref{SimTable}. 

For all the models, including the standard joint model, 
fixed effects for both the longitudinal and survival 
models and baseline hazard parameters demonstrate similar 
biases and coverages. In terms of averages of the 
credibility intervals for $\alpha_1$ and $\alpha_2$, 
the standard joint model produces larger values compared to 
the other models.  Coverages for the elements 
of ${\bf \Sigma}$ are lower than expected 
for the standard joint model, 
and almost 0 for Approach 1, whereas 
they are at the expected level for Approaches 2, 3 and 4.
Approaches 1 and 2 produces biased and 0 coverage 
results for $\sigma^2$. 
This highlights the need to remove the dependence between $Z_{ij}$ and $Z_{ij'}$.   
Approach 3 provides similar estimates and levels of coverage compared to Approach 4. Therefore, whilst Approach 3 has the inability to fully capture the time-varying nature of the degrees-of-freedom for the residuals, this appears to have limited impact on the estimation of the other parameters. 

 \section{Illustrative Data Examples}
 \label{DataAnalysis}

\subsection{Renal Example}
\label{DataAnalysis:renal}
The first case-study considers NI haemodialysis patients' data with a joint model specified as:
\begin{eqnarray}
Y_{ij} &=& Y_i(t_{ij}) + Z_{ij},\nonumber\\
       &=& \alpha_{1} + t_{ij}\alpha_{2} +  \text{age}_{i}\alpha_{3}  + \text{creatinine}_{ij}\alpha_{4} + \text{MCHC}_{ij}\alpha_{5} + \text{MCV}_{ij}\alpha_{6} + \text{urea}_{ij}\alpha_{7}  \nonumber \\ 
&& + \text{male}_{i}\alpha_{8} + \text{I(EPO}_{ij} \text{=Aranesp)}\alpha_{9} + \text{I(EPO}_{ij} \text{=Alfa)}\alpha_{10} + \text{I(EPO}_{ij} \text{=Other)}\alpha_{11} \nonumber \\
&& + \text{Prior2007}_{i}\alpha_{12} + \text{IronHydroxide}_{ij}\alpha_{13} + \text{Venofer}_{ij}\alpha_{14}  + B_{1i} + t_{ij}B_{2i} + Z_{ij} \\
h_i(t) & =& h_0(t)  \exp \left( \text{age}_{i}\omega + Y_i^*(t) \eta \right),
\end{eqnarray}
where the response $Y$ is an individual's Hb level, 
$h_{0}(t)$ is assumed to be Weibull baseline hazard, 
MCHC represents mean corpuscular haemoglobin concentration, MCV represents mean corpuscular volume and EPO represents erythropoietin drug treatments.

The models were fitted using the {\tt robjm} package with four chains each of length 8,000 with the first 2,000 iterations considered as warm-up. Posterior summaries for each of the approaches are given within Table~\ref{table:joint:renal}. Convergence was confirmed using traceplots and R-hat statistics~\citep{brooks97}.

\begin{figure}[ht!]
\centering
  \includegraphics[scale = 0.7]{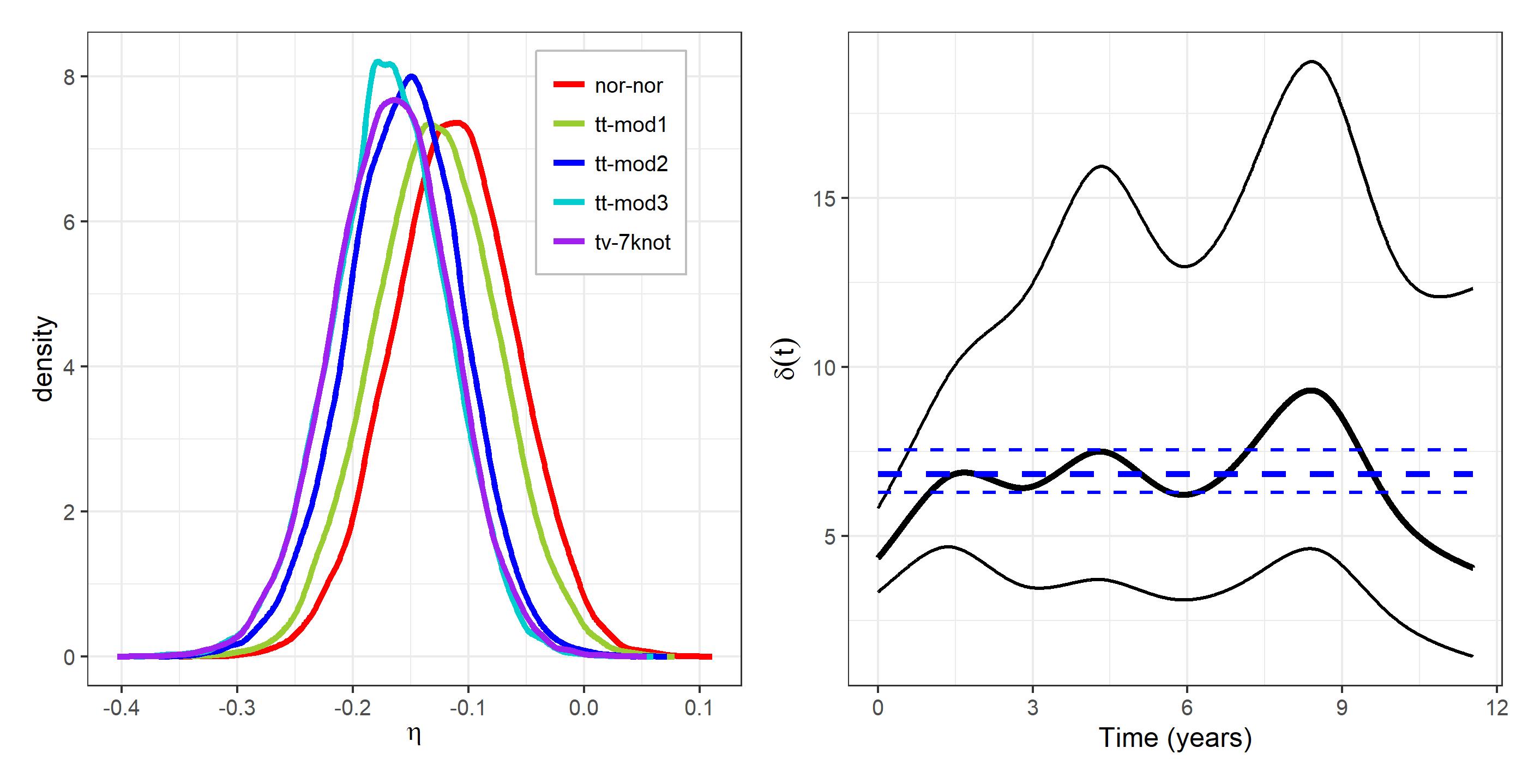}
  \caption{Left panel: Posterior densities of $\eta$ 
  under different models. Right panel: Medians of 
  degrees-of-freedom parameters (mid-lines) 
  and 2.5\% and 97.5\% percentiles of the MCMC samples 
under Approach 3 (solid) and time-varying model with 7 knots (dashed) based on NI renal data.}
  \label{DOFPlot_renal}
\end{figure}

Similar to the findings of the simulation study in Section~\ref{SimulationStudy}, in general, results based on Approaches 
3 and 4 are quite similar. There are differences in the results of  
these two models and the others, some parameters 
were overestimated whereas some were underestimated. For example,  
the association parameter, $\eta$, appears to be affected by the choice of model. It is underestimated by the 
standard joint model, which was overcome by Approach 1 up to some 
extent. Smoothed density plots for $\eta$ 
are displayed in the left panel of Figure~\ref{DOFPlot_renal}. 
Posterior summaries of degrees-of-freedom for 
Approach 3 and Approach 4 with 7 knots 
are displayed in the right panel 
of Figure~\ref{DOFPlot_renal}.  

\subsection{PBC Example}
 \label{sec:pbcexample}
 
 The second case-study uses data from the PBC study.  
The model we fit has the following form:
\begin{align}
Y_{ij} &= Y_i^*(t_{ij}) + Z_{ij} \nonumber\\ 
       &= \alpha_{1} + \text{age}_i \alpha_2 + t_{ij} \alpha_3 + \text{I(drug$_i$ = D-penicillamine)} \alpha_4 + B_{i1} + t_{ij} B_{i2} + Z_{ij},\\
 h_i(t) &= h_0(t) \exp \left(\text{age}_i \omega_1 +  
 \text{I(drug$_i$ = D-penicillamine)} \omega_{2} + 
 Y_i^*(t) \eta \right), 
\end{align}
where $Y = \log(\text{alkaline})$,
age is baseline age, $t$ follow-up time, 
$h_0(t)$ Weibull baseline 
hazard function. 

For Approaches 2--4, the results for $\phi$ indicated that 
Normal assumption might be reasonable for ${\boldsymbol{\mathrm{B}}}_i$.
Therefore, only $Z$ term was assumed to be $t$-distributed. 
For Approach 4,  
we considered number of knots ranging from 1 to 5. 
The knots were put into the empirical quantiles of the 
time variable. Four chains, each length of 6,000, were started from 
random initials. Half of each chain was considered 
as warm-up. Convergence was checked through 
traceplots and R-hat statistics \citep{brooks97}. 
Posterior summaries, specifically the 
2.5\%, 50\% and 97.5\% percentiles, are 
displayed in Table \ref{tab:pbc_results}.  
 
\begin{figure}[!ht]
\centering
  \includegraphics[scale = 0.65]{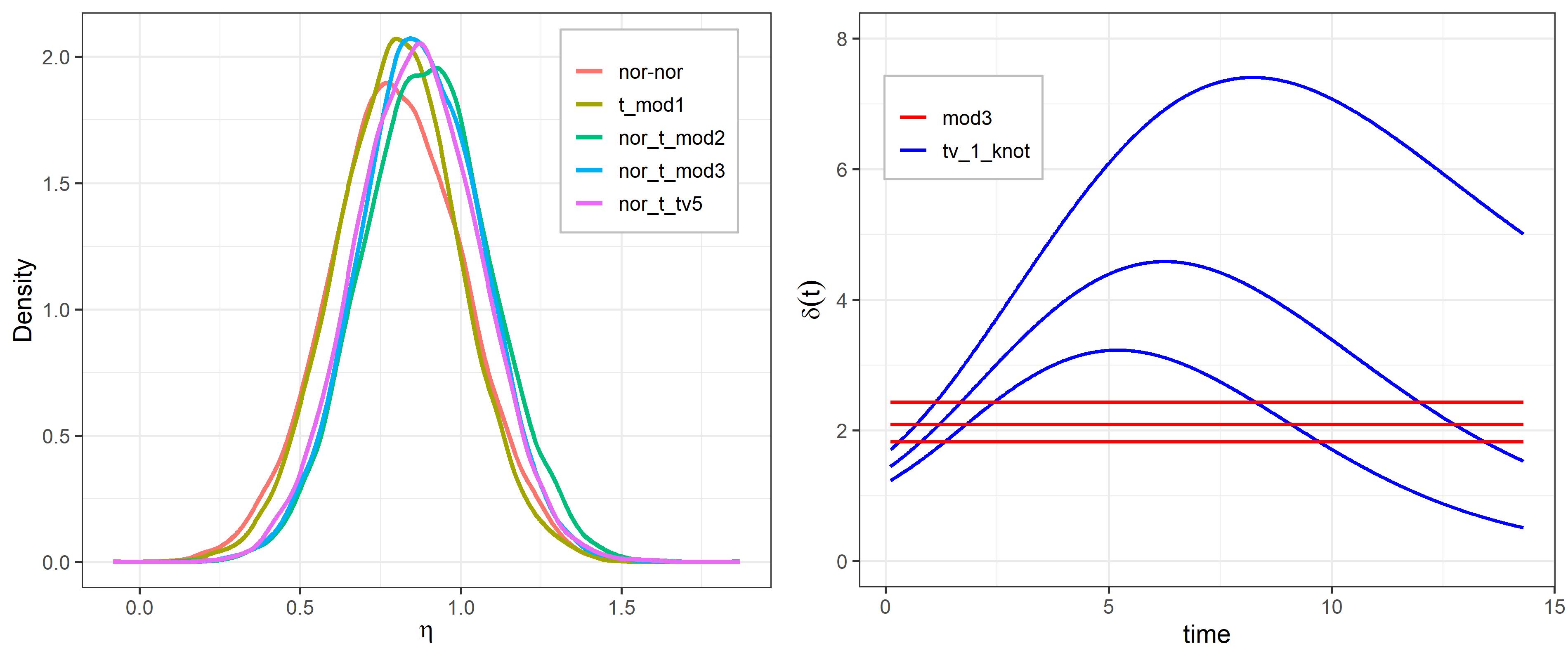}
  \caption{Left panel: smoothed MCMC samples of $\eta$ 
  under different models. 
  Right panel: Medians of 
  degrees-of-freedom parameters (mid-lines) 
  and 2.5\% and 97.5\% percentiles of 
  the MCMC samples under model 3 and tv model with 1 knot.}
  \label{fig:pbc_eta_dof}
\end{figure}

The association parameter, $\eta$, 
seems to be affected by the model choice. 
Regarding model comparisons in terms of parameter estimates, 
similar observations to those made for the NI renal application  were observed for this case-study.  
The smoothed densities of the MCMC samples of $\eta$ 
and  the degrees-of-freedom estimates for Approach 
3 and Approach 4  
with 1 knot are displayed in the left and right panels 
of Figure~\ref{fig:pbc_eta_dof}, 
respectively.

{\tt R} codes to run the models for the PBC data 
could be reach at {\tt \url{https://github.com/ozgurasarstat/robjm/blob/master/tests/testthat/test_pbc_analysis.R}}.

\section{Discussion}
\label{Discussion}

This work introduces a new flexible approach to model 
longitudinal outliers in a joint modelling framework, 
where the degrees-of-freedom parameter for the residuals 
is assumed to vary over time. This scenario replicates the common situation when patients take time to adjust to new treatments, resulting in more outlying measurements (e-outliers) at different times across the period of observation. In addition, this paper contrasts the new time-varying approach with three alternative time-invariant formulations of robust joint models currently found in the literature and the standard joint model with Gaussian assumptions for the random terms. Each of the approaches presented can be fitted in the accompanying {\tt robjm} software package.

Throughout the results presented within the simulation study and the real-life data analysis, the need to properly account for longitudinal outliers is evident, a practice not widely adopted in the current joint modelling literature. This paper highlights the need to investigate the presence of outliers and provides a flexible approach to downweighing their negative impact, contrasting the effect of various assumptions made by the different approaches to robust joint modelling. It is evident that bias and inefficiency results from ignorance of longitudinal outliers and the assumption of normality for the random terms. In addition, this work highlights the need to remove the restrictive assumptions of Approaches 1 and 2. 

\begin{figure}[!ht]
\centering
  \includegraphics[scale = 0.7]{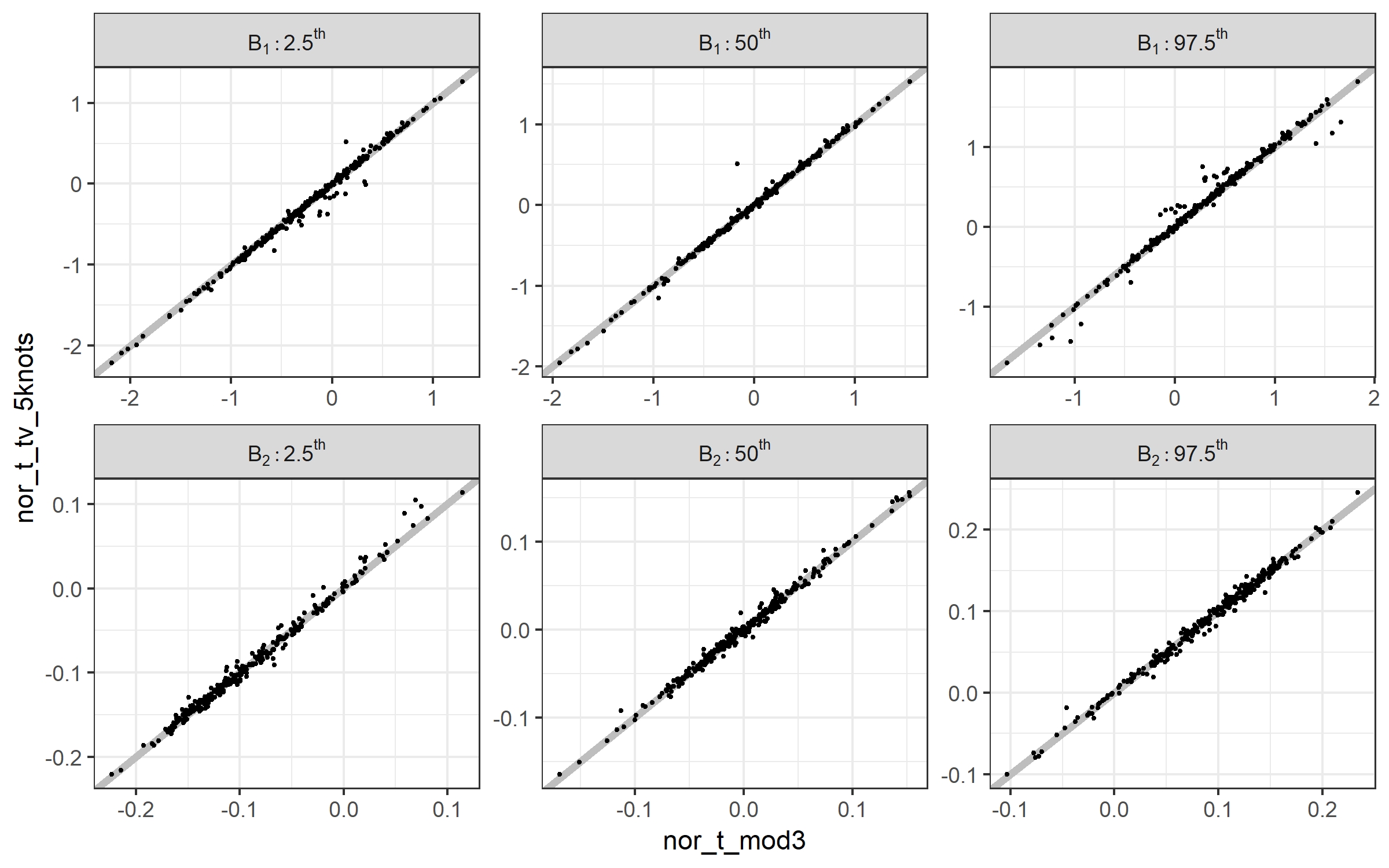}
  \caption{Posterior summaries (2.5\%, 50\% and 97.5\% percentiles of 
  the MCMC samples) for Approach 3 against the time-varying model with 5 knots from the PBC analysis for the predictions of the longitudinal random intercept (top row) and slope (bottom row). }
  \label{fig:Bplot_nor_t_mod3_nor_t_tv_5knots}
\end{figure}

Unexpectedly, despite a true underlying assumption of time-varying degrees-of-freedom within the simulation study, Approach 3 was found to be in good agreement with the time-varying model. Within the simulation study and in the two real-life examples, the degrees-of-freedom were assumed or estimated to be consistently low and varying over a limited range throughout the period of observation. Under such scenarios, Approach 3 appears to provide a time-invariant average of the time-varying degrees-of-freedom for the residuals over the period of observation, as demonstrated within Figure~\ref{DOFPlot_renal} for the renal analysis. Given the limited change in estimated parameters between the two approaches, it is not therefore unexpected to find high levels of agreement when comparing the posterior summaries of the predictions for the longitudinal random effects, as illustrated in Figure~\ref{fig:Bplot_nor_t_mod3_nor_t_tv_5knots} from the PBC analysis.

This work highlights the robustness provided by Approach 3 in the estimation of the degrees-of-freedom, providing similar estimates as the time-varying approach inspite of the presence of time-varying degrees-of-freedom. It signifies that in cases where the degrees-of-freedom parameter does not vary widely, Approach 3 is recommended when longitudinal outliers are present.

We did not consider model selection 
methods to select the number of knots, mainly because the 
number of knots does not seem to change the results considerably. 
In the time-varying degrees-of-freedom model we assumed the scale parameter, 
$\sigma$ being time-constant. This parameter might also be assumed 
to be time-varying.

\section*{Funding}
This work was supported by the Engineering and Physical Sciences Research Council [Reference: EP/P026028/1] 

\subsection*{Acknowledgment}

Dr. Laura Boyle's (The University of Adelaide) helps 
are gratefully acknowledged. 

%\subsection*{Financial disclosure}
%
%None reported.

\subsection*{Conflict of interest}

The authors declare no potential conflict of interests.

%\vspace{0.5cm}

\clearpage

\vspace{-0.1cm}

\begin{sidewaystable}
\centering
\caption{Table showing the simulation results based on data assuming time-varying degrees-of-freedom}
\label{SimTable}\vspace{0.4cm}
\hspace*{-0.5cm}\begin{tabular}{ccl r ccc ccc ccc ccc ccc}
\cline{3-19}
&& \multicolumn{1}{c}{} &  & \multicolumn{3}{c}{Standard Joint} & \multicolumn{3}{c}{Approach 1} & \multicolumn{3}{c}{Approach 2} & \multicolumn{3}{c}{Approach 3} & \multicolumn{3}{c}{Approach 4} \\
\cline{5-19}
 &&& \multicolumn{1}{c}{True} & \multicolumn{1}{c}{Mean} &  \multicolumn{1}{c}{CI} &  \multicolumn{1}{c}{Cov.} &  \multicolumn{1}{c}{Mean} &  \multicolumn{1}{c}{CI} &  \multicolumn{1}{c}{Cov.} & \multicolumn{1}{c}{Mean} &  \multicolumn{1}{c}{CI} &  \multicolumn{1}{c}{Cov.} & \multicolumn{1}{c}{Mean} &  \multicolumn{1}{c}{CI} &  \multicolumn{1}{c}{Cov.} & \multicolumn{1}{c}{Mean} &  \multicolumn{1}{c}{CI} &  \multicolumn{1}{c}{Cov.} \\
\cline{3-19}
&&&&& &&&&& &&&&& &&& \vspace{-0.3cm}\\								
&	&	$\alpha_1$	&	1.00	& 0.992	&	0.329	& 0.930		&	0.986	&	0.291	& 0.935	&	0.994	& 0.247    & 0.920   & 0.995   & 0.247   & 0.910    & 0.995  & 0.246 & 0.920 	\\
	&	&	$\alpha_2$	&	0.40 	&	0.391	& 	0.233	& 0.940		&	0.389	&	0.209	&	0.950	&	0.396	& 0.175    & 0.935   &  0.396  & 0.176   & 0.930    & 0.396	 & 0.175 & 0.930	\\
	&	&	${\boldsymbol{\Sigma}}(1,1)^{*}$	& \	\ 0.60	&	1.703	&	0.638 & 0.495		&	1.234	&	0.523	&	0.005	&	0.622	& 0.377    & 0.955   &   0.619 & 0.373   & 0.970    &	0.616 & 0.372 & 0.955 \\
	&	&	${\boldsymbol{\Sigma}}(2,1)^{*}$	& \	0.25	&	0.673	& 0.363		& 0.590		&	0.497	&	0.287	&	0.070	&	0.251	& 0.184    & 0.960   &  0.251  & 0.183   & 0.955    & 0.252	 & 0.183  & 0.955 \\
	&	&	${\boldsymbol{\Sigma}}(2,2)^{*}$	& \ 0.30	&	0.823	&	0.328 	&	0.490	&	0.608	&	0.265	&	0.000	&	0.311	& 0.188    & 0.960   & 0.311   & 0.188   & 0.965    &	0.310 & 0.187 & 0.965 \\
	&	&	${\gamma}^{**}$	&	---	&	---	&	---	&	---	&	20.719	 &	---	&	---	&	---	& ---    & ---   &  ---  & ---   & ---    &	---  & --- & --- \\
    &	&	$\phi$	&	3.00	&	---	&	---	&	---	&	---	 &	---	&	---	&	3.169	& 2.285    & 0.930   &  3.163  & 2.286   & 0.940    &	3.159  & 2.270 & 0.940 \\
	&	&	${\sigma^2}^{***}$	&	0.25	&	0.351	&	---	&	---	&	0.322	&	0.044	&	0.000	&	0.325	& 0.043    & 0.000   &  0.250  & 0.042   & 0.965    & 0.259	 & 0.044  & 0.825  \\
	&	&	${\delta}^{***}$	&	5.00	&	---	&	---	&	---	& ---		&	---	&	---	&	30.350	&    --- & ---  & 7.149   & ---   & ---    & ---	 & ---  & --- \\
    &	&	$\delta_0$	&	5.00	&	---	&	---	&	---	& ---		&	---	&	---	&	---	&    --- & ---  & ---   & ---   & ---    & 5.801	 & 6.160  & 0.950 \\
		&	& $\beta_1$	&	0.50	&	---	&	---	&	---	& ---		&	---	&	---	& ---	& ---	& ---	& ---	& ---	& ---	& 0.536 & 2.285   & 0.960  \\
		&	& $\beta_2$	&	0.50	&	---	&	---	&	---	& ---		&	---	&	---	& ---	& ---	& ---	& ---	& ---	& ---	&0.668 & 2.648 & 0.945  \\
		&	& $\beta_3$	&	0.50	&   ---	&	---	&	---	& ---		&	---	&	---	& ---	& ---	& ---	& ---	& ---	& ---	&0.612 & 2.508 & 0.935  \\
		&	& $\beta_4$	&	0.50	& ---	&	---	&	---	& ---		&	---	&	---	& ---	& ---	& ---	& ---	& ---	& ---	&0.614	 & 2.41  & 0.935   \\
		&	& $\beta_5$	&	0.50	&	---	&	---	&	---	& ---		&	---	&	---	& ---	& ---	& ---	& ---	& ---	& ---	&0.693 & 2.921  & 0.940   \\
		&	& $\beta_6$	&	-0.25	&	---	&	---	&	---	& ---		&	---	&	---	& ---	& ---	& ---	& ---	& ---	& ---	&0.152 & 2.400  & 0.895  \\	
	&	&	$\lambda$	&	0.04	&	0.040	&	0.036	&	0.950	&	0.040	&	0.036	&	0.955	&	0.040	& 0.036    & 0.950   & 0.040   & 0.036   & 0.945    &	 0.040 & 0.036 & 0.950 \\
	&	&	$\nu$	&	1.20	& 1.207		& 0.433 		&	0.950	&	1.206	&	0.434	&	0.955	&	1.210	& 0.430    & 0.945   & 1.210   & 0.431   & 0.950    & 1.211	 & 0.432  & 0.950  \\
	&	&	$\omega$	&	0.50 	&	0.494	& 0.709		&	0.945	&	0.494	&	0.709	& 0.945   	&	0.494	& 0.708    & 0.940   & 0.494   & 0.709   & 0.935    &	0.494 & 0.713 & 0.950 \\		
		&	&	$\eta$	&	0.30 	&	0.308	&	0.139	&	0.955	&	0.309	&	0.140	&	0.945	&	0.306	& 0.137    & 0.960   &  0.307  & 0.138   & 0.950    &	0.307 & 0.138 & 0.950 \\		
\cline{3-19}
\multicolumn{19}{l}{} \vspace{-0.1cm} \\		
\multicolumn{19}{l}{\hspace{0.5cm} $^*$ \hspace{0.02cm} For standard joint model, true values for these parameters were considered as 1.8 $\left(0.6 * \frac{3}{3-2}\right)$, 0.75 $\left(0.3 * \frac{3}{3-2}\right)$ and 0.9 $\left(0.3 * \frac{3}{3-2}\right)$.}\\
\multicolumn{19}{l}{\hspace{0.5cm} $^{**}$ \hspace{0.02cm} For this parameter, CI and Cov. were not calculated, since the truth 
is unknown.}\\
\multicolumn{19}{l}{\hspace{0.5cm} $^{***}$ \hspace{0.02cm} For these parameters, CI and Cov. were not calculated, since the truths  are time-varying.}
\end{tabular}
\end{sidewaystable}

%\begin{landscape} 
%\begin{sidewaystable}[t]
\begin{table}{t}
\vspace{-2cm}  
\centering
\caption{Posterior summaries of the joint model parameters fitted to the NI renal dataset. Values in the upper rows for each parameter are the 50\% percentiles, whereas the brackets in the lower rows are the 2.5\% and 97.5\% percentiles, respectively. }
\label{table:joint:renal} 
\scalebox{0.6}{
\begin{tabular}{l c c c c c c c c c c}\hline
 & \multicolumn{1}{c}{nor-nor} & \multicolumn{1}{c}{tt-mod1} & \multicolumn{1}{c}{tt-mod2} & \multicolumn{1}{c}{tt-mod3} & \multicolumn{1}{c}{tv-2knot} & 
   \multicolumn{1}{c}{tv-3knot} & \multicolumn{1}{c}{tv-4knot} &\multicolumn{1}{c}{tv-5knot} & \multicolumn{1}{c}{tv-6knot} & \multicolumn{1}{c}{tv-7knot}\\ \hline
\multirow{2}{*}{$\alpha_1$} &  10.138        &   10.239         &    10.261       &    10.246       &     10.245       &     10.248       &    10.247        &    10.251        &  10.249 & 10.250 \\  
                         & (10.018, 10.259)   & (10.117, 10.354)  & (10.145, 10.376) & (10.133, 10.361) & (10.130, 10.357)  & (10.132, 10.360)  & (10.134, 10.360)  & (10.135, 10.364) & (10.132, 10.364) & (10.136, 10.365) \\            
\multirow{2}{*}{$\alpha_2$} &  -0.133        &  -0.138         &    -0.130       &     -0.127       &   -0.126         &     -0.127       &           -0.126 &  -0.126           &  -0.126 &-0.126  \\  
                            & (-0.163, -0.103)  & (-0.167, -0.110)  & (-0.157, -0.103) & (-0.154, -0.100) & (-0.154, -0.099)  & (-0.154, -0.099) & (-0.154, -0.099) & (-0.154, -0.099) & (-0.154, -0.099) & (-0.154, -0.099) \\   
\multirow{2}{*}{$\alpha_3$} &   0.124       &    0.112       &    0.104        &    0.104         &  0.105           &    0.105         &          0.105   &     0.105        &  0.105   & 0.105  \\  
                            & (0.089, 0.159)  &  (0.077, 0.147) & (0.071, 0.136)  & (0.073, 0.137) & (0.073, 0.138)  & (0.073, 0.137) & (0.073, 0.137) & (0.072, 0.137) & (0.073, 0.138)  & (0.073, 0.138) \\   
\multirow{2}{*}{$\alpha_4$} &    0.138      &   0.125        &      0.123     &   0.127          &    0.128         &    0.128         &          0.128   &    0.128         & 0.128    & 0.128  \\  
                            & (0.121, 0.154) &  (0.109, 0.142) & (0.107, 0.140) & (0.111, 0.144) & (0.111, 0.145) & (0.111, 0.144)  & (0.112, 0.144) & (0.111, 0.144) & (0.111, 0.144)  & (0.111, 0.145) \\   
\multirow{2}{*}{$\alpha_5$} &  0.257        &   0.259        &       0.257    &    0.259         &  0.259           &       0.259      &          0.259   &   0.259          &  0.259   & 0.258  \\  
                            & (0.239, 0.275) & (0.242, 0.276) & (0.240, 0.274) & (0.241, 0.276) & (0.241, 0.276)  & (0.241, 0.276) & (0.241, 0.276) & (0.241, 0.276)  & (0.241, 0.276)  & (0.241, 0.276) \\   
\multirow{2}{*}{$\alpha_6$} &   0.265       &   0.290        &    0.281       &    0.274         &      0.277       &     0.277        &          0.277   &     0.277        &  0.276   & 0.276   \\  
                            & (0.222, 0.308) & (0.245, 0.334) &  (0.238, 0.325) & (0.231, 0.318) & (0.234, 0.320)  & (0.234, 0.321)  & (0.233, 0.319) & (0.233, 0.320)  & (0.233, 0.320)  & (0.232, 0.320) \\   
\multirow{2}{*}{$\alpha_7$} &   0.097       &    0.102       &     0.101       &   0.109          &    0.110         &     0.109       &          0.110   &   0.110          &  0.109   & 0.109  \\  
                            & (0.075, 0.120)  &  (0.080, 0.125) & (0.079, 0.123) & (0.087, 0.132) & (0.087, 0.132) & (0.087, 0.132)  & (0.087, 0.132) & (0.087, 0.132) & (0.087, 0.131)  & (0.087, 0.132) \\   
\multirow{2}{*}{$\alpha_8$} &   -0.155       &   -0.142        &     -0.178       &    -0.187        &      -0.187      &    -0.190        &           -0.190  &    -0.191        &  -0.191  & -0.191 \\  
                            & (-0.255, -0.055) & (-0.238, -0.045) & (-0.270, -0.082) & (-0.281, -0.095) & (-0.279, -0.095)  & (-0.280, -0.100)  & (-0.280, -0.097)  & (-0.282, -0.098) & (-0.282, -0.099)  & (-0.282, -0.100) \\   
\multirow{2}{*}{$\alpha_9$} &  0.207        &     0.187      &   0.176           &   0.195          &     0.193        &    0.193         &          0.192   &   0.192          &  0.191   & 0.191  \\  
                            & (0.136, 0.278) & (0.117, 0.257)  &  (0.107, 0.243) & (0.125, 0.261) & (0.124, 0.261) & (0.125, 0.263)  & (0.123, 0.261) & (0.124, 0.258) & (0.123, 0.261)  & (0.122, 0.258) \\   
\multirow{2}{*}{$\alpha_{10}$} &   -0.037       &  -0.062         &    -0.064       &    -0.053        &   -0.051         &  -0.051          &   -0.050         &    -0.051        &  -0.052  & -0.052 \\  
                            & (-0.147, 0.074)  & (-0.171, 0.045) & (-0.170, 0.041) & (-0.161, 0.054) & (-0.160, 0.054)  & (-0.158, 0.055)  & (-0.157, 0.057)  & (-0.157, 0.054)  & (-0.160, 0.055)  & (-0.158, 0.057) \\   
\multirow{2}{*}{$\alpha_{11}$} &    -0.100      &  -0.140         &  -0.152          &   -0.119         &    -0.131        &   -0.130         & -0.130           &   -0.135         &  -0.139  & -0.140 \\  
                            &  (-0.342, 0.151) & (-0.381, 0.099)  & (-0.385, 0.089) & (-0.346, 0.101) & (-0.357, 0.092) & (-0.357, 0.096)  & (-0.356, 0.093)  & (-0.364, 0.088)  & (-0.366, 0.087)  & (-0.364, 0.089) \\   
\multirow{2}{*}{$\alpha_{12}$} &  0.348         &  0.333         &     0.331       &    0.325         &    0.324         &     0.324        &    0.324         &    0.322         &  0.323   & 0.323  \\  
                            & (0.246, 0.451) & (0.236, 0.433) & (0.239, 0.426) & (0.231, 0.419) & (0.230, 0.417) & (0.231, 0.418)  & (0.228, 0.418)  & (0.206, 0.419) & (0.228, 0.417)  & (0.230, 0.419) \\   
\multirow{2}{*}{$\alpha_{13}$} &    0.329      &   0.292        &    0.282         &    0.310         &    0.310         &     0.310        &   0.311          &    0.311         &  0.311   & 0.311  \\  
                            & (0.271, 0.388)  & (0.236, 0.348)  & (0.227, 0.337) & (0.255, 0.367) & (0.254, 0.366)  & (0.254, 0.367)  & (0.254, 0.367) & (0.254, 0.367)  & (0.255, 0.367)  & (0.254, 0.368) \\   
\multirow{2}{*}{$\alpha_{14}$} &   0.241       &    0.223       &      0.218      &   0.221          &     0.221        &    0.221         &           0.221   &    0.220         &  0.220  & 0.220  \\  
                            &  (0.201, 0.282) & (0.185, 0.261) & (0.180, 0.255) & (0.182, 0.260) & (0.182, 0.259) & (0.182, 0.260) & (0.182, 0.260)  & (0.179, 0.259) & (0.181, 0.259)  & (0.182, 0.259) \\   
\multirow{2}{*}{$\Sigma_{11}$} &   1.043       &    0.901       &  0.551          &    0.543         &    0.540         &     0.538        &     0.539        &     0.534        &  0.536   & 0.536  \\  
                            &  (0.940, 1.158) & (0.806, 1.005) & (0.452, 0.664)  & (0.452, 0.648) & (0.448, 0.644) & (0.445, 0.640)  & (0.446, 0.644) & (0.442, 0.638) & (0.447, 0.638)  & (0.444, 0.643) \\   
\multirow{2}{*}{$\Sigma_{12}$} &  -0.256      &   -0.218        &    -0.136       &   -0.140         &     -0.138      &    -0.137        &   -0.137         &    -0.136        &  -0.137  & -0.137 \\  
                            & (-0.303, -0.213) & (-0.258, -0.182) & (-0.172, -0.108) & (-0.175, -0.110) & (-0.173, -0.108)  &  (-0.171, -0.107) & (-0.173, -0.108)  & (-0.171, -0.107) & (-0.170, -0.108)  & (-0.172, -0.107) \\   
\multirow{2}{*}{$\Sigma_{22}$} &   0.134       &   0.113        &     0.075       &    0.080         &    0.078         &      0.078       &   0.077          &    0.076         &  0.077   & 0.077   \\  
                            & (0.111, 0.162) & (0.094, 0.138)  & (0.059, 0.095) & (0.063, 0.101) & (0.061, 0.098) & (0.061, 0.097) & (0.061, 0.098) & (0.060, 0.097) & (0.061, 0.097)  & (0.061, 0.098) \\   
\multirow{2}{*}{$\sigma^2$} &   1.260       &    1.010       &   0.969         &    0.894         &    0.894         &  0.894           &          0.895   &     0.897        &  0.895   & 0.897  \\  
                            & (1.238, 1.283)  & (0.972, 1.048) & (0.931, 1.008) & (0.866, 0.925) & (0.865, 0.925)  & (0.866, 0.925) & (0.866, 0.926) & (0.868, 0.925) & (0.866, 0.926)  & (0.866, 0.928) \\   
\multirow{2}{*}{$\gamma$} &   \multirow{2}{*}{---}        &  8.685         &  \multirow{2}{*}{---}      &      \multirow{2}{*}{---}       &     \multirow{2}{*}{---}        &     \multirow{2}{*}{---}        &  \multirow{2}{*}{---}           &     \multirow{2}{*}{---}        &  \multirow{2}{*}{---}   & \multirow{2}{*}{---}  \\  
                            &  & (7.708, 9.802) &  &  &   &   &   &  &   &  \\
\multirow{2}{*}{$\phi$} &   \multirow{2}{*}{---}        &  \multirow{2}{*}{---}         &  3.150      &      2.764       &     2.760        &     2.742        &  2.751           &     2.727        &  2.726   & 2.727  \\  
                            &  &  & (2.504, 4.132) & (2.280, 3.419) & (2.273, 3.407)  & (2.263, 3.382)  & (2.283, 3.417)  & (2.244, 3.364) & (2.261, 3.370)  & (2.256, 3.392) \\    
\multirow{2}{*}{$\delta$} &  \multirow{2}{*}{---}        & \multirow{2}{*}{---}         &       7.789      &     6.847         &   \multirow{2}{*}{---}          &     \multirow{2}{*}{---}        &    \multirow{2}{*}{---}         &       \multirow{2}{*}{---}      &  \multirow{2}{*}{---}   & \multirow{2}{*}{---}  \\  
                            &  & & (6.908, 8.804)  & (6.296, 7.561) &   & &   &  &  &  \\
\multirow{2}{*}{$\delta_0$} &  \multirow{2}{*}{---}        & \multirow{2}{*}{---}         &       \multirow{2}{*}{---}      &     \multirow{2}{*}{---}         &   5.532          &     5.071        &    4.671         &       4.894      &  4.554   & 4.355  \\  
                            &  & &   &  & (4.722, 6.600)  & (4.167, 6.250) & (3.782, 5.911)  & (3.874, 6.338) & (3.553, 6.016) & (3.347, 5.833) \\
\multirow{2}{*}{$\beta_1$} &  \multirow{2}{*}{---}        & \multirow{2}{*}{---}         &    \multirow{2}{*}{---}        & \multirow{2}{*}{---}         &          0.609        &    0.277         &    0.212        &     0.501         &    0.336       &        0.261               \\  
                            & & & & & (0.272, 0.957)  & (0.041, 0.513)  & (-0.056, 0.489) & (0.182, 0.833)  & (-0.032, 0.695) &  (-0.122, 0.652)   \\   
\multirow{2}{*}{$\beta_2$} &  \multirow{2}{*}{---}        & \multirow{2}{*}{---}         &    \multirow{2}{*}{---}        & \multirow{2}{*}{---}         &  0.089       &   0.704          &    0.589         &       0.044      &      0.546      &      0.694                    \\  
                            & & & & & (-0.346, 0.529)  & (0.333, 1.097)  & (0.258, 0.921) & (-0.348, 0.435) & (0.106, 0.984) &  (0.234, 1.178)  \\   
\multirow{2}{*}{$\beta_3$} &  \multirow{2}{*}{---}        & \multirow{2}{*}{---}         &    \multirow{2}{*}{---}        & \multirow{2}{*}{---}         &   -0.865      &    0.281        &     0.432        &    0.781          &   0.129        &      0.233                 \\  
                           & & & & & (-1.363, -0.328) & (-0.249, 0.830)  & (0.006, 0.858) & (0.447, 1.115) & (-0.263, 0.513)  &  (-0.181, 0.667)   \\   
\multirow{2}{*}{$\beta_4$} &  \multirow{2}{*}{---}        & \multirow{2}{*}{---}         &    \multirow{2}{*}{---}        & \multirow{2}{*}{---}         &  \multirow{2}{*}{---}      &   -0.991         &     0.468       &       0.165       &     0.983      &                   0.542   \\  
                            & & & & & & (-1.548, -0.379) & (-0.165, 1.074)  & (-0.288, 0.619)  & (0.600, 1.396) & (0.086, 0.984)   \\   
\multirow{2}{*}{$\beta_5$} &  \multirow{2}{*}{---}        & \multirow{2}{*}{---}         &    \multirow{2}{*}{---}        & \multirow{2}{*}{---}         &  \multirow{2}{*}{---}      &   \multirow{2}{*}{---}        &    -0.812        &     0.198        &   -0.049        &        0.939                 \\  
                            &  &  & & & & & (-1.407, -0.162) & (-0.483, 0.879) & (-0.548, 0.459) &  (0.526 ,1.340)   \\   
\multirow{2}{*}{$\beta_6$} &  \multirow{2}{*}{---}        & \multirow{2}{*}{---}         &    \multirow{2}{*}{---}        & \multirow{2}{*}{---}         &  \multirow{2}{*}{---}      &    \multirow{2}{*}{---}       &     \multirow{2}{*}{---}       &          -0.526   &   0.399         &        -0.004                \\  
                            &  &  &  & & & & & (-1.156, 0.204)  & (-0.317, 1.132) & (-0.516, 0.499)   \\   
\multirow{2}{*}{$\beta_7$} &  \multirow{2}{*}{---}        & \multirow{2}{*}{---}         &    \multirow{2}{*}{---}        & \multirow{2}{*}{---}         &  \multirow{2}{*}{---}      &    \multirow{2}{*}{---}       &     \multirow{2}{*}{---}       &          \multirow{2}{*}{---}  &       -0.358     &         0.524                \\  
                            &  &  &  & & & & & & (-1.052, 0.442)  &  (-0.230, 1.263)   \\ 
\multirow{2}{*}{$\beta_8$} &   \multirow{2}{*}{---}        & \multirow{2}{*}{---}         &    \multirow{2}{*}{---}        & \multirow{2}{*}{---}         & \multirow{2}{*}{---}      &    \multirow{2}{*}{---}       &     \multirow{2}{*}{---}       &          \multirow{2}{*}{---}  &     \multirow{2}{*}{---}       &       -0.411        \\  
                            &  &  &  & &  & & & & & (-1.138, 0.390)    \\ 
\multirow{2}{*}{$\lambda$}  &  0.261        &    0.314         &    0.415          &     0.482        &    0.467         &   0.479          &          0.476   &      0.484       &  0.480   & 0.476  \\  
                            & (0.079, 0.868)  & (0.095, 1.030)  & (0.140, 1.312) & (0.164, 1.505) & (0.160, 1.508) & (0.166, 1.524)  & (0.159, 1.499) & (0.163, 1.565) & (0.159, 1.536)  & (0.153, 1.512) \\   
\multirow{2}{*}{$\nu$} &  1.278        &   1.277         &     1.272        &     1.272        &      1.273       &      1.273       &          1.273   &     1.272        &  1.273   & 1.273  \\  
                            & (1.195, 1.363)  & (1.194, 1.361) & (1.187, 1.356) & (1.191, 1.357) & (1.192, 1.358) & (1.191, 1.357)  & (1.191, 1.357) & (1.191, 1.356) & (1.190, 1.358)  & (1.192, 1.356) \\   
\multirow{2}{*}{$\omega$}&   0.268       &     0.269        &     0.273        &   0.274          &  0.273           &    0.273         &          0.273   &    0.274         &  0.273   & 0.273  \\  
                            &  (0.203, 0.334) & (0.203, 0.336) & (0.209, 0.339)  & (0.209, 0.339) & (0.209, 0.340)  & (0.210, 0.340) & (0.209, 0.339) & (0.209, 0.339) & (0.208, 0.341)  & (0.208, 0.340) \\   
\multirow{2}{*}{$\eta$} &  -0.113        &     -0.129      &    -0.154       &     -0.168       &       -0.165     &   -0.167         &          -0.166  &    -0.168        &  -0.167  & -0.166 \\  
                            & (-0.219, -0.009) & (-0.234, -0.024)  & (-0.257, -0.059) & (-0.270, -0.073) & (-0.269, -0.071)  & (-0.270, -0.073)  & (-0.268, -0.070)  & (-0.273, -0.073)  & (-0.271, -0.070)  & (-0.270, -0.068) \\  
  \hline
\end{tabular}}
%\end{sidewaystable}
\end{table}

%\end{landscape}

%\begin{landscape} 
 
 \begin{sidewaystable}[t]
 \centering
\caption{Posterior summaries of the joint model parameters fitted to the PBC dataset. Values in the upper rows for each 
parameter are the 50\% percentiles, whereas the 
brackets in the lower rows are the 2.5\% and 97.5\% 
percentiles, respectively. }
\label{tab:pbc_results} 
\scalebox{0.65}{
\begin{tabular}{l c c c c c c c c c}\hline
 & \multicolumn{1}{c}{nor-nor} & \multicolumn{1}{c}{mod1} & \multicolumn{1}{c}{nor-t-mod2} & \multicolumn{1}{c}{nor-t-mod3} & \multicolumn{1}{c}{nor-t-tv-1knot} & 
   \multicolumn{1}{c}{nor-t-tv-2knots} & \multicolumn{1}{c}{nor-t-tv-3knots} &\multicolumn{1}{c}{nor-t-tv-4knots} & \multicolumn{1}{c}{nor-t-tv-5knots} \\ \hline
\multirow{2}{*}{$\alpha_1$} & 7.851          & 7.468          & 7.479          & 7.515          & 7.521          & 7.517          & 7.521          & 7.526          & 7.529 \\  
                            & (7.259, 7.907) & (7.159, 7.779) & (7.161, 7.792) & (7.205, 7.840) & (7.191, 7.848) & (7.194, 7.832) & (7.206, 7.856) & (7.210, 7.863) & (7.197, 7.842)\\            
\multirow{2}{*}{$\alpha_2$} &  -0.008          &   -0.007         &    -0.006       &   -0.007         &  -0.007           & -0.007           &  -0.007          &  -0.007          &  -0.007 \\  
                            & (-0.014, -0.001) & (-0.013, -0.001) & (-0.012, 0.001) & (-0.013, -0.001) & (-0.014, -0.0003) & (-0.013, -0.001) & (-0.014, -0.001) & (-0.014, -0.001) & (-0.013, -0.001)\\
\multirow{2}{*}{$\alpha_3$} &  -0.046          &  -0.041          &  -0.040          &  -0.035          &  -0.035          &  -0.035          &     -0.035       &  -0.035          &  -0.035\\  
                            & (-0.058, -0.034) & (-0.051, -0.030) & (-0.052, -0.028) & (-0.046, -0.024) & (-0.046, -0.024) & (-0.046, -0.023) & (-0.046, -0.024) & (-0.046, -0.023) & (-0.046, -0.023)\\
\multirow{2}{*}{$\alpha_4$} &  -0.066         &   -0.020        &  -0.059         &  -0.072         &  -0.073         &     -0.071      &   -0.072        &    -0.070       &  -0.073 \\  
                            & (-0.185, 0.054) & (-0.133, 0.095) & (-0.184, 0.062) & (-0.199, 0.051) & (-0.200, 0.049) & (-0.193, 0.054) & (-0.198, 0.050) & (-0.192, 0.051) & (-0.192, 0.046)\\
\multirow{2}{*}{$\Sigma_{11}$} &   0.339       &  0.230         &   0.320        &    0.349       &  0.357         &   0.359        &     0.360      &  0.359         & 0.358 \\  
                              & (0.286, 0.406) & (0.186, 0.284) & (0.268, 0.384) & (0.295, 0.414) & (0.301, 0.425) & (0.304, 0.426) & (0.303, 0.428) & (0.303, 0.428) & (0.302, 0.427)\\
\multirow{2}{*}{$\Sigma_{12}$} &   -0.018         &   -0.009         &     -0.012       &    -0.015        &   -0.016         &     -0.016       &   -0.016         &    -0.016        &  -0.016 \\  
                               & (-0.026, -0.011) & (-0.015, -0.004) & (-0.019, -0.005) & (-0.022, -0.009) & (-0.023, -0.009) & (-0.023, -0.009) & (-0.024, -0.010) & (-0.023, -0.010) & (-0.024, -0.010)\\      
\multirow{2}{*}{$\Sigma_{22}$} &    0.004       &  0.003         &    0.004       &  0.005         &   0.005        &  0.005         &  0.005         &   0.005        &  0.005\\  
                               & (0.003, 0.005) & (0.002, 0.004) & (0.003, 0.006) & (0.004, 0.006) & (0.004, 0.006) & (0.004, 0.006) & (0.004, 0.006) & (0.004, 0.006) & (0.004, 0.006)\\ 
\multirow{2}{*}{$\sigma^2$} &   0.107        &   0.047        &    0.040       &  0.021         &   0.023        &   0.024        &  0.024         &  0.024         & 0.024 \\  
                            & (0.099, 0.115) & (0.040, 0.054) & (0.034, 0.047) & (0.018, 0.025) & (0.020, 0.027) & (0.021, 0.028) & (0.021, 0.028) & (0.021, 0.029) & (0.021, 0.029)\\
\multirow{2}{*}{$\gamma$} &  \multirow{2}{*}{---}    &   3.516        &  \multirow{2}{*}{---}        &   \multirow{2}{*}{---}        &   \multirow{2}{*}{---}        &  \multirow{2}{*}{---}         &     \multirow{2}{*}{---}      &    \multirow{2}{*}{---}       &  \multirow{2}{*}{---} \\  
                                        &      & (2.810, 4.394) &  &  &  &  &  &  & \\
\multirow{2}{*}{$\delta$} &  \multirow{2}{*}{---}    &   \multirow{2}{*}{---}        &  2.759         &   2.091        &  \multirow{2}{*}{---}        &   \multirow{2}{*}{---}         &     \multirow{2}{*}{---}      &   \multirow{2}{*}{---}       &   \multirow{2}{*}{---} \\  
                                        &      &  & (2.207, 3.451) & (1.824, 2.432) &  &  &  &  & \\
\multirow{2}{*}{$\delta_0$} &  \multirow{2}{*}{---}    &   \multirow{2}{*}{---}        &    \multirow{2}{*}{---}        &   \multirow{2}{*}{---}        &   1.449        &  1.184         &     1.130      &    1.122       &  1.130 \\  
                                        &      &  &  &  & (1.235, 1.704) & (0.994, 1.410) & (0.940, 1.361) & (0.931, 1.360) & (0.935, 1.367)\\
\multirow{2}{*}{$\beta_1$} & \multirow{2}{*}{---}     &  \multirow{2}{*}{---}    &  \multirow{2}{*}{---}    &  \multirow{2}{*}{---}    &   2.001        &    -0.527       &  1.358         &  1.308         & 1.432 \\  
                           &      &      &      &      & (1.432, 2.625) & (-1.408, 0.339) & (0.782, 2.009) & (0.591, 2.118) & (0.664, 2.035)\\
\multirow{2}{*}{$\beta_2$} &  \multirow{2}{*}{---}    & \multirow{2}{*}{---}     &  \multirow{2}{*}{---}    & \multirow{2}{*}{---}     &  -0.543        &  2.946         &  0.202          &   1.376        &  1.175 \\  
                           & 		  & 		 & 		  & 		 & (-1.459, 0.459) & (2.076, 3.969) & (-0.673, 1.031) & (0.699, 2.128) & (0.391, 2.032)\\
\multirow{2}{*}{$\beta_3$} &  \multirow{2}{*}{---}    &   \multirow{2}{*}{---}   &   \multirow{2}{*}{---}   &   \multirow{2}{*}{---}   &   \multirow{2}{*}{---}   &  1.380         &  2.566         &  0.242          & 1.329 \\  
                           & 		  & 		 & 		  & 		 & 		  & (0.079, 3.056) & (1.720, 3.549) & (-0.816, 1.241) & (0.694, 2.024)\\
\multirow{2}{*}{$\beta_4$} &  \multirow{2}{*}{---}    &  \multirow{2}{*}{---}    &   \multirow{2}{*}{---}   &   \multirow{2}{*}{---}   &   \multirow{2}{*}{---}   &   \multirow{2}{*}{---}   &   1.210         &  2.414         &  0.314 \\  
                           & 		  &      & 		  & 		 & 		  & 		 & (-0.192, 3.202) & (1.516, 3.694) & (-0.703, 1.352)\\
\multirow{2}{*}{$\beta_5$} &  \multirow{2}{*}{---}    &   \multirow{2}{*}{---}   &   \multirow{2}{*}{---}   &   \multirow{2}{*}{---}   &   \multirow{2}{*}{---}   &   \multirow{2}{*}{---}   &   \multirow{2}{*}{---}   &  1.763         &  2.263\\  
                           & 		  &      &      & 		 & 		  & 		 & 		  & (0.107, 4.256) & (1.321, 3.556)\\
\multirow{2}{*}{$\beta_6$}  &   \multirow{2}{*}{---}   &   \multirow{2}{*}{---}   &   \multirow{2}{*}{---}   &   \multirow{2}{*}{---}   &   \multirow{2}{*}{---}   &   \multirow{2}{*}{---}   &   \multirow{2}{*}{---}   &    \multirow{2}{*}{---}  &  1.909\\  
                            &      &      & 		 & 		  & 		 & 		  &      &      & (0.161, 4.574)\\
\multirow{2}{*}{$\log(\lambda)$}  &  -11.276          &   -11.427         &    -12.129        &   -11.952         &   -11.901         &    -11.911        &   -11.861          &     -11.854        & -11.872 \\  
                                  & (-14.850, -7.844) & (-14.814, -8.168) & (-15.584, -8.869) & (-15.215, -8.761) & (-15.300, -8.641) & (-15.256, -8.773) & (-15.178, -8.694) & (-15.200, -8.618) & (-15.242, -8.679)\\
\multirow{2}{*}{$\log(\nu)$} &   0.223        &   0.226        &   0.234        &   0.222        &  0.221         &   0.220        &  0.219         &  0.220         & 0.221 \\  
                             & (0.064, 0.373) & (0.068, 0.372) & (0.076, 0.380) & (0.069, 0.366) & (0.065, 0.367) & (0.062, 0.367) & (0.065, 0.364) & (0.062, 0.362) & (0.066, 0.363)\\
\multirow{2}{*}{$\omega_1$}&     0.051      &   0.052        &    0.053       &   0.053        &    0.053       &     0.053      &  0.053         &    0.053       &  0.053\\  
                           & (0.033, 0.070) & (0.035, 0.070) & (0.036, 0.072) & (0.035, 0.071) & (0.035, 0.072) & (0.036, 0.072) & (0.035, 0.071) & (0.036, 0.071) & (0.035, 0.072)\\
\multirow{2}{*}{$\omega_2$} &  -0.157         &   -0.121        &   -0.144        &    -0.137       &      -0.137     &    -0.136       &     -0.136      &     -0.133      & -0.139 \\  
                            & (-0.519, 0.206) & (-0.460, 0.229) & (-0.508, 0.212) & (-0.488, 0.217) & (-0.492, 0.220) & (-0.498, 0.228) & (-0.494, 0.215) & (-0.483, 0.217) & (-0.490, 0.215)\\
\multirow{2}{*}{$\eta$} &    0.799       &   0.805        &   0.895        &   0.873        &   0.872        &     0.869      &    0.865       &   0.862        &  0.865 \\  
                        & (0.391, 1.216) & (0.423, 1.193) & (0.511, 1.293) & (0.502, 1.244) & (0.491, 1.253) & (0.497, 1.255) & (0.487, 1.245) & (0.482, 1.250) & (0.488, 1.248)\\ \hline
\end{tabular}}
\end{sidewaystable}

%\end{landscape}

\end{document}